\documentclass[preprint,review,12pt]{elsarticle}
\usepackage[toc,page]{appendix} 
\usepackage{fancyhdr}
\usepackage{verbatim}
\usepackage{xcolor}
\usepackage[us,12hr]{datetime} % `us' makes \today behave as usual in TeX/LaTeX
\fancypagestyle{plain}{
\fancyhf{}
\rfoot{P.\thepage}

}

\usepackage{graphicx}
\usepackage{amssymb}
\usepackage{amsmath}
\newcommand{\ar}{\text{Ar}}
\newcommand{\xe}{\text{Xe}}

\journal{Nuclear Instruments and Methods in Physics Research Section A}

\begin{document}
\begin{frontmatter}


 \title{Understanding the Enhancement of Scintillation Light in Xenon-Doped Liquid Argon}
 %{\it \small version {\ddmmyyyydate\today} at \currenttime} }

 %{\it version {\ddmmyyyydate\today} at \currenttime} }




\author[UNM]{D.E.~Fields}
\author[UNM,current]{R.Gibbons}
\author[UNM]{M.~Gold}
\author[UNM]{N.~McFadden}
\author[LANL]{S.R.~Elliott}
\author[LANL]{R.~Massarczyk}

\address[UNM]{Department of Physics and Astronomy
MSC07 4220, 1 University of New Mexico, Albuquerque NM 87131-0001}
\address[LANL]{Physics Division, Los Alamos National Laboratory MS H803, P-23, Los Alamos, NM, 87545, USA}
\address[current]{currently at Physics Department, Bldg. 50, Lawrence Berkeley National Laboratory, Berkeley, California, 94720-50}

\begin{abstract}
Measuring the scintillation light in noble gases is an important detection technique in particle physics. Numerous rare event searches like neutrino beam experiments,
neutrino-less double beta-decay, and dark matter searches use argon-based detectors. In liquid argon, the 
light yield can be enhanced by the addition of a small quantity of xenon, where $\sim 10 - 1000$ ppm are added. The general enhancement mechanism and its pathway via an energy transfer between argon and xenon excimers is well known, however
 the importance of absorption of argon excimer emission by atomic xenon has not been fully appreciated. This absorption significantly reduces the light yield in commercially available argon (extracted from air) which contains trace amounts ($\rm \sim 0.1$ ppm) of xenon.  The addition of a small xenon dopant of $\sim 10$ ppm recovers this lost light resulting in an increased light yield over un-doped argon of about a factor of two.
 In this paper we introduce a 
model for the light production in xenon doped argon, including absorption and re-emission, and compare it
to the measured time dependence of light emission in xenon-doped argon.

\end{abstract}


\end{frontmatter}

%\linenumbers


%%\cite{mcfadden2020largescale} 
%%\cite{Kubota1993}
%%\cite{Neumeier2015}

\section{Introduction}
\label{S:Intro}
The use of liquid noble gases in detectors in physics experiments has become a mainstay. In particular, liquid argon (LAr) has become widely used because of its low cost. A number of experiments use scintillation light in low rate, large volume/mass applications needed in dark sector searches\cite{Ajaj:2019imk},\cite{CCM},\cite{Agnes_2018} or neutrino detection\cite{PhysRevLett.126.012002}, or as active shield detector for neutrino-less double beta decay experiments like GERDA\cite{Agostini_2020} or LEGEND\cite{LEGENDOverview}. Due to the short wavelength of the 
argon scintillation light ($128$ nm) and the low sensitivity of common light detectors at this range, LAr experiments generally require the use of a wavelength shifting material.
Common materials like Tetraphenyl Butadiene, (TPB), have to be coated on detectors, light guides, or on the walls of the cryostats. The possibility of shifting the light within the LAr itself to a different wavelength is therefore intriguing. One possibility is the doping of LAr with xenon with the subsequent creation of xenon excimers which emit light at higher wavelength ($175$ nm)\cite{Peiffer:2008zz}. Previous works have demonstrated that a small dopant concentration of $\sim10$ ppm is sufficient to transfer 
a large portion of the argon scintillation to the xenon emission wavelength\cite{Neumeier2015},\cite{McFadden:2020dxs}. 

When energy is deposited in pure LAr, argon excimers form which emit light at 128 nm.
It is known that trace amounts ($\sim 0.1$ ppm) of xenon remain in commercially available argon extracted from air, leading to absorption of the argon excimer light\cite{Neumeier:2015lka}.  In this paper we present a model for this absorption that agrees with our measurement of the light emission as a function of time. This model explains changes in light yield as a function of xenon concentration in the LAr.

\section{Light emission in xenon-doped argon} 
\begin{figure}
\begin{center}
 \includegraphics[width=0.9 \columnwidth]{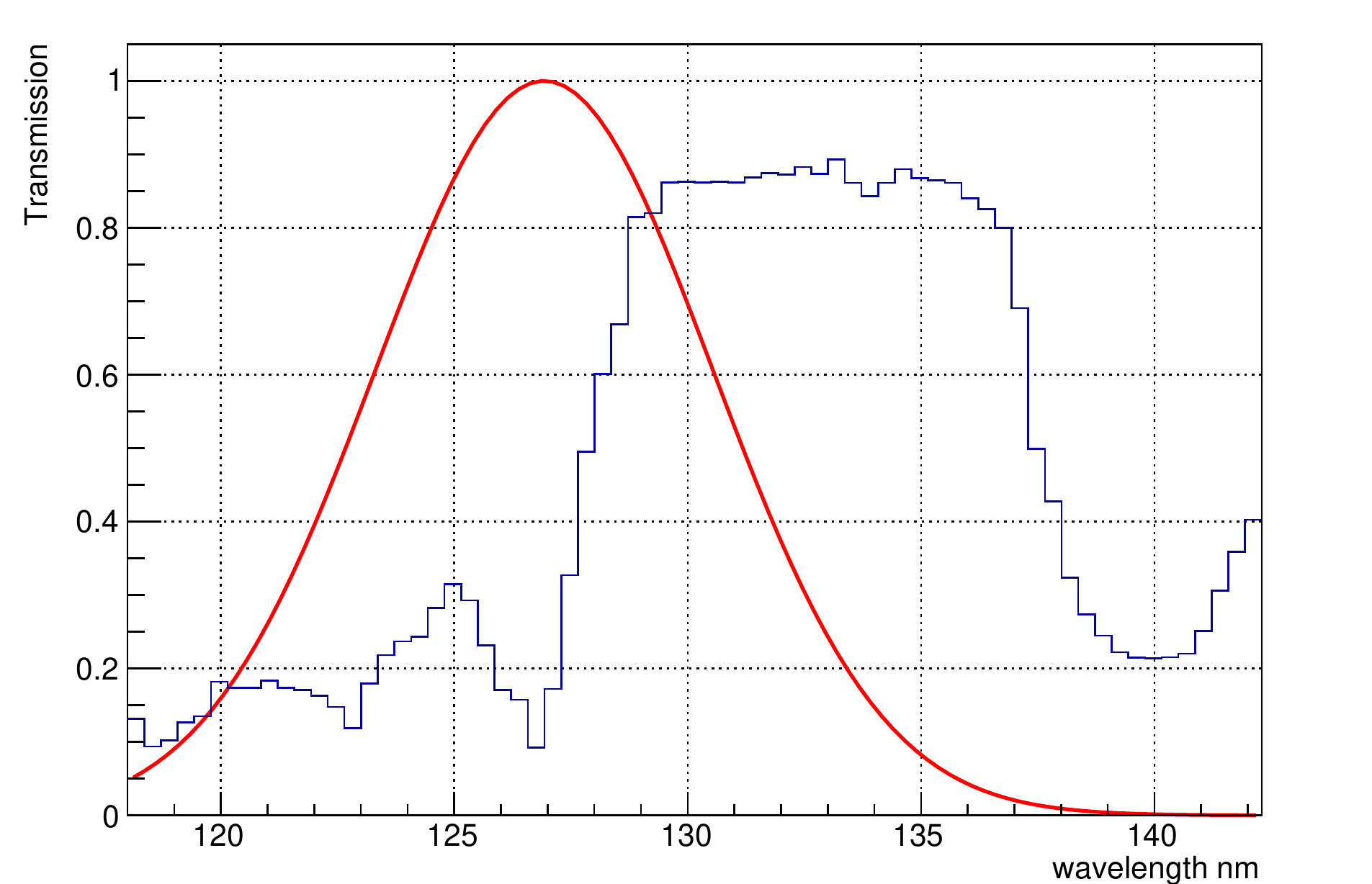}
 \caption{Transmission of light as a function of wavelength through 11.6 cm path of liquid argon with 0.1 ppm xenon\cite{Neumeier:2015lka}. Superimposed (red line) is the argon excimer 128 nm emission line.
}
\label{fig-transDoped}
\end{center}
\end{figure}

The existence of trapped exciton states of xenon in liquid argon was discovered by Raz and Jortner\cite{RazJortner1970}. These states were observed in the absorption spectrum of liquid argon with $\sim 0.1$ ppm xenon\cite{Neumeier:2015lka}. The measured absorption of light through a path length of 11.6 cm liquid argon is shown in figure~\ref{fig-transDoped} together with the 128 nm argon excimer emission line superimposed.

\begin{figure}
\begin{center}
 \includegraphics[width=0.9 \columnwidth]{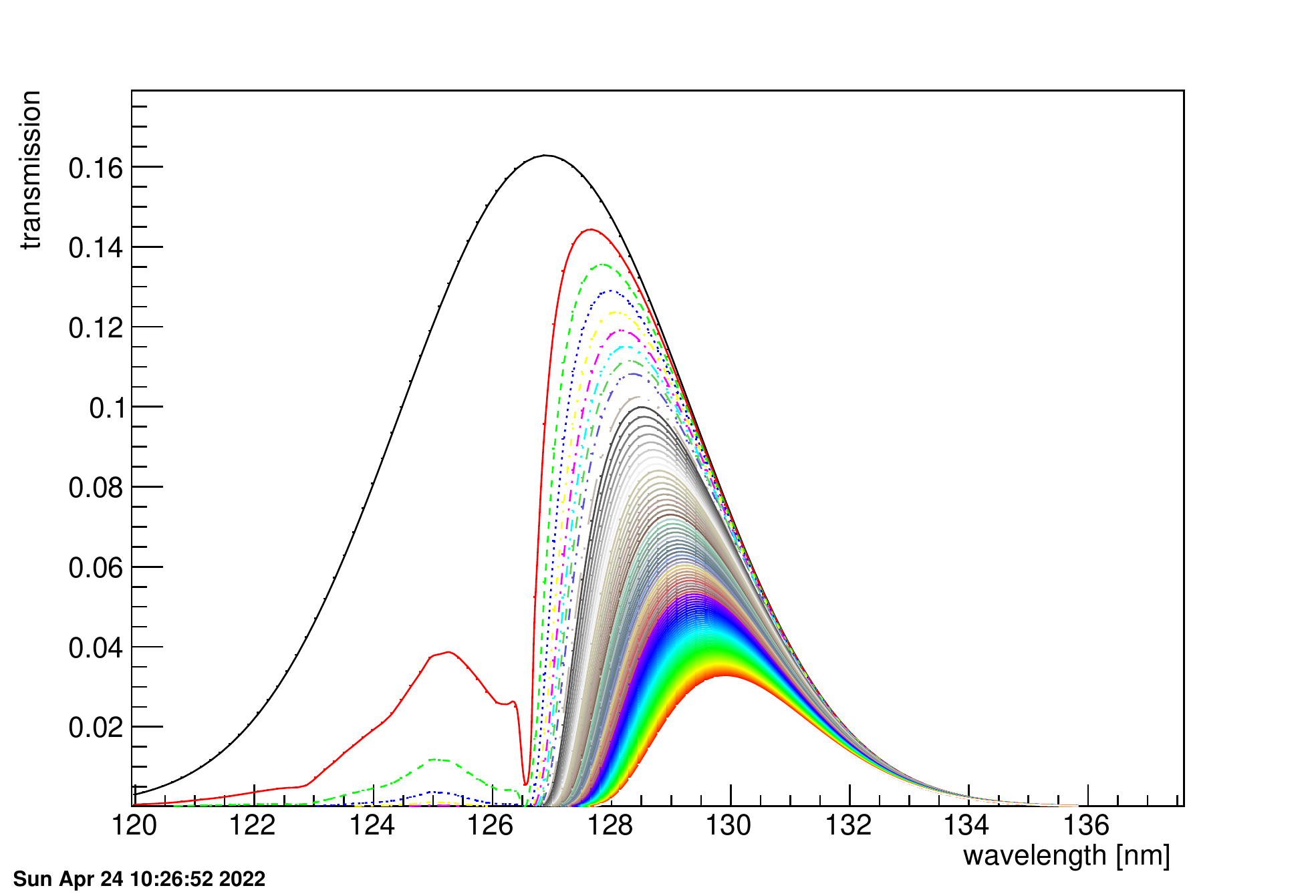}
 \caption{Results of the convolution of the two graphs of Fig. \ref{fig-transDoped}. Each curve is an additional step of 11.6 cm (the initial path length) over 100 steps with a maximum distance of 1160 cm.
}
\label{fig-transByStep}
\end{center}
\end{figure}

This absorption spectrum is convoluted with the argon excimer emission to obtain the results seen in figure~ \ref{fig-transByStep}, where it is clear that even at low levels of xenon (0.1 ppm) and relatively short distances (11.6 cm), more than half of the initial 128 nm light is absorbed.

\begin{figure}[htbp]
\begin{center}
 \includegraphics[width=0.9 \columnwidth]{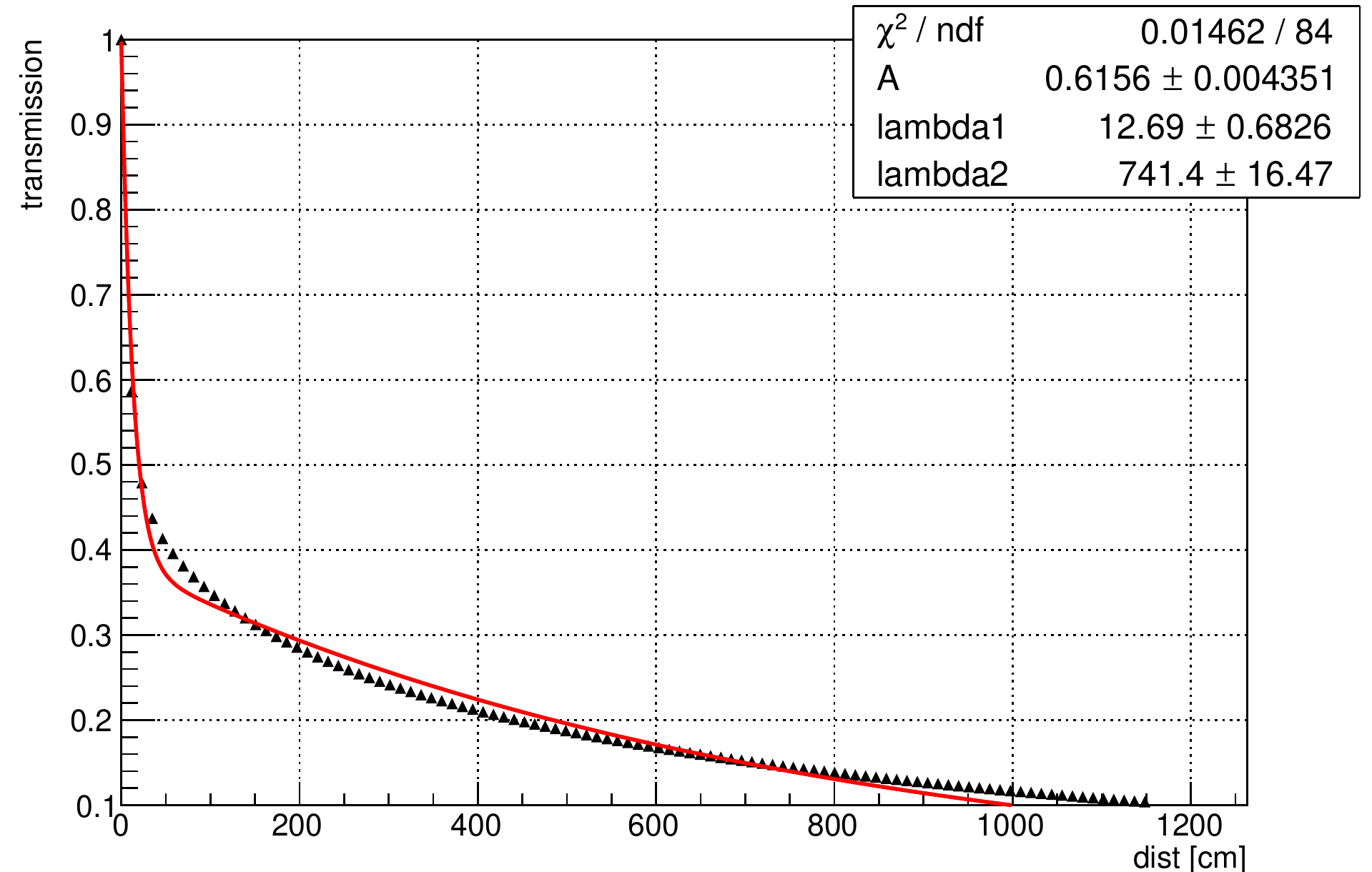}
 \caption{Transmission of argon 128 nm light with distance from convolution of argon emission spectrum with transmission data of figure~\ref{fig-transDoped}. The red curve is a double exponential fit to the data. 
}
\label{fig-attVdist}
\end{center}
\end{figure}

These curves are then integrated over wavelength to obtain the transmission as a function of distance shown in  figure~\ref{fig-attVdist}. This transmission curve is then fitted to the double exponential:  
\begin{equation}
T=Ae^{-x/\lambda_1}+(1-A)e^{-x/\lambda_2}.
\end{equation}
From the fit, for distances in the range from 10-100 cm, the effect is to absorb an approximately constant factor of $A=0.62$ of the total light, taken from the fit values.  This is a key parameter in our model.  The cross section for this absorption in liquid argon can be written as $\sigma_{abs} = 1/nl_{abs}c$, where n is the atomic number density of liquid argon ($2.1 \times 10^{28}/m^3$), $l_{abs}$ is the effective absorption length from the fit ($12.7 cm$, see fit value of $\lambda_1$ in figure~\ref{fig-attVdist}), and c is the concentration of xenon atoms in parts per million (0.1 ppm, as given in the reference for the xenon concentration for their data). The result is $\sigma_{abs}=37.5Mb$, in agreement with the result of Ref.~\cite{CALVO2018186}. It should be noted that the horizontal axis of the curve (and thus the fit parameters $\lambda_1$ and $\lambda_2$) of figure~\ref{fig-attVdist} can be scaled for different xenon concentrations by the factor (xenon concentration in ppm)/(0.1 ppm).

%Ref. \cite{Segreto:2020qks} describes in detail the creation of light in argon via excitons and electron-hole pairs. 
Argon excimers $\ar_2^*$ are created in singlet ($S$) or triplet ($T$) states with about 85\% in the T state for gammas~\cite{Segreto:2020qks}~\footnote{The ratio of singlet to triplet states depends on the type of ionizing radiation, presence of an electric field, etc.}. The argon excimer singlet $S$ has a lifetime of $\rm 5 \; ns$, whereas the triplet $T$ has a much longer lifetime of $\rm \sim 1600 \; ns$. As these excimers decay, xenon atoms absorb light and form an exciton state.  This state
immediately forms the mixed state $\ar\xe^*$\cite{Kubota1993} excimer. The lifetime of the mixed state is long, $\rm 4700 \; ns$, and has a decay wavelength of 150 nm\cite{Soto-Oton:2021rjm}.  There is a competing process of collisional quenching of the excimer states at a rate of 
about $\rm 1/(7.7 \mu s)$\cite{Hitachi} that contributes to total light loss.  The mixed state can form a xenon excimer $\xe_2^{*}$ through diffusion to xenon at a rate which increases with the xenon dopant. 

%As discussed in Ref. , possible ionized dimers can quickly neutralized to $\ar_2^*$. Therefore, we focus in our model on the interaction of the excited dimers. While both states, $S$ and $T$, emit light with almost the same wavelength, the singlet state is very short-lived while the triplet lives longer.
%In the presence of xenon, the argon dimers can transfer energy to the xenon atoms. This process shifting energy from argon dimer $\ar_2^*$ to xenon dimer $\xe_2^*$ is well understood to proceed by the formation of an intermediate mixed state $\ar\xe^*$\cite{Hitachi,Buzulutskov:2017wau}. The lifetime of the mixed state is long, $\rm 4700 \; ns$ \cite{Soto-Oton:2021rjm}. As a reference, the argon singlet $S$ has a lifetime of $\rm 5 \; ns$, the triplet $T$ has a lifetime of $\rm \sim 1600 \; ns$, and the xenon dimer $\xe_2^*$ has a lifetime of $\rm 20 \; ns$. 
%It has been shown that the lifetime of $T$ strongly dependant on purity and absorption/re-emission \cite{1989JChPh..91.1469M}. 
%The peak wavelength of the emissions are argon 128 nm, mixed 150 nm and xenon 175 nm. 

 Our model consists of four coupled differential equations for the number of molecules as a function of time $t$ for the excimer states argon singlet $S(t)$, argon triplet $T(t)$, and the combined singlet and triplet states of the mixed $M(t)$ and xenon $X(t)$:
 
 \begin{equation}
 \dot{S}=-S/\tau_{S} -\left( k_x+k_q \right) S \equiv -\lambda_1 S
 \label{eqone}
 \end{equation}
  \begin{equation}
 \dot{T}=-T/\tau_T-\left( k_x+k_q \right) T \equiv -\lambda_3 T 
 \label{eqtwo}
 \end{equation}
 \begin{equation}
 \dot{M}= -\left(1/\tau_M + k_x + k_q^\prime \right) M + \left( k_x+A/\tau_{S} \right) S + \left( k_x+A/\tau_{T} \right) T 
 \label{eqthree}
 \end{equation}
 \begin{equation}
 \dot{X} = - X/\tau_x + k_x M 
 \label{eqfour}
 \end{equation}
 In these equations, the lifetimes of the different states are labeled using $\tau$ in combination with the corresponding letter as index. We introduce collisional de-excitation factors due to quenching $k_q = 1.3 \times 10^{-4}\;ns^{-1}$, and $k^\prime_q \sim k_q $, as used in Ref.~\cite{Hitachi}, although we note that the collisional quenching may be stronger for the more weakly bound M and X excimers.
 
 %The xenon dopant reaction rate $k_x$ can be calculated from the diffusion-limited reaction rate in liquid argon as
%\begin{equation}
%k_{x} = 4\pi D_{AB}R_{AB}C
%\end{equation}
%where $D_{AB}$ is the self diffusion constant of liquid argon, $R_{AB}$ is the effective reaction distance between an argon eximer and a Xe atom, and C is the concentration of xenon atoms.  If one uses $\rm D_{AB} = 1.53 \times 10^{-4}cm^2/s$ \cite{CiniCastognoli} and $\rm R_{AB} = 4.84 \times 10^{-8}cm$ (using atomic Van der Waals radii ) gives $\rm k_{x} = 2.4 \times 10^{-4}[ppm] \; ns^{-1}$. Wahl, et. al, uses a value of $\rm k_x  = 2.9\times 10^{-4}[ppm] \; ns^{-1} $\cite{CGWahl} without reference, with the implication of a larger effective interaction radius.  In this work, we adopt the larger value, but note here that we plan to determine the effective radius in a future work with finer dopant increments. 
We use $\rm k_x  = 2.9\times 10^{-4}[ppm] \; ns^{-1} $\cite{CGWahl}.\footnote{This number appears without reference in~\cite{CALVO2018186}. It is close to the value of $\rm k_x  = 2.4\times 10^{-4}[ppm] \; ns^{-1} $ calculated from the diffusion limited reaction rate using
Van der Walls values for the atomic radii\cite{CiniCastognoli}. }
The absorption of the argon excimer deexcitation light at 128nm is parameterized by the dimensionless constant $A$ which we take from the fit in figure~\ref{fig-attVdist}. Although this number varies with dopant, especially at dopant levels less than 1 ppm, it is unknown what the xenon concentration was in the undoped LAr of our experiment, therefore we use the same absorption across all data sets.
 
  Integrating these equations, the total emitted light yield $\ell(t)$ as a function of time is given by,
\begin{equation}
\ell(t) = N_1 \left( 1-A \right) e^{-t \lambda_1}/\tau_S + N_3 \left( 1-A \right) e^{-t\lambda_3} /\tau_T+ M(t)/\tau_M+ X(t)/\tau_X
\end{equation}
 where $N_1$ and $N_3$ are the number of initial singlet and triplet states, respectively. The decay constants are defined using the introduced half-lifes $\tau$: $\lambda_X = 1/\tau_X$, $k_x^\prime = 1/\tau_M + k_x + k_q^\prime$, $C_1= k_x + A/\tau_S$, $C_3= k_x + A/\tau_T$. The total number of mixed states as a function of time $ M(t) = M_1(t)+ M_3(t)$ is 
 
\begin{equation}
M_i(t) = \frac{N_i C_i} {\lambda_i - k_x^\prime} \left[ e^{-t k_x^\prime } - e^{-t \lambda_i } \right]
\end{equation}
for $i=1,3$.

Similarly, the number of xenon states can be described as $X(t)= X_1(t) + X_3(t) $ with

\begin{equation} X_i(t) = \frac{N_i C_i k_x} {\lambda_i - k_x^\prime} \left[ \frac{ e^{-t k_x^\prime } - e^{-t \lambda_X } } {\lambda_x - k_x^\prime } 
											 - \frac{ e^{-t \lambda_i } - e^{-t \lambda_X } } {\lambda_x - \lambda_i} \right],
\end{equation}
again for $i=1,3$.
When fitting this distribution to the data, one has to consider that the timing resolution of our detector smears out the exact distributions. We used the measured rise time of the singlet light emission (see figure~\ref{fig-singlet-summed} to determine the timing resolution of our setup, and fold this as a Gaussian distribution into the exponential decay. 

\section{Experimental Setup}

The liquid argon test stand at UNM is described in detail in Ref.~\cite{McFadden:2020dxs}. The same reference shows that the measured properties of the argon are in good agreement with previous measurements, and that the doping system is well understood.
%demonstrated by introducing a nitrogen contamination instead of xenon. 
The apparatus is a 100L cylindrical liquid argon cryostat with a single PMT mounted on the bottom facing upwards. The 3$''$ Hamamatsu R11065 Photomultiplier (PMT) has a TPB coated acrylic disk fixed to the front of the tube to shift the light to a wavelength with good detection efficiency.\footnote{It should be noted here that the wavelength dependence of TPB is essentially flat over the region 120 - 200 nm\cite{TPB_efficiency}, so no additional correction is needed for this study} The cryostat was filled with argon gas and liquefied using a CryoMech cold head. The amount of LAr in the cryostat was monitored by constantly measuring the weight of the cryostat.  On top of the liquid phase, a gas phase is present due to boil-off.  The argon gas was circulated through a SAES PS4-MT3/15-R getter, which purifies noble gasses to less than 1 PPB of: $H_20$, $CO$, $CO_2$, $N_2$, $H_2$, $CH_4$
prior to liquefaction\cite{Getter}. The operation of the getter and the doping system was verified previously with the introduction of a known amount of nitrogen and monitoring the light yield initially decreasing and then recovering\cite{McFadden:2020dxs}. For the data sets presented in this work, xenon dopant was added to achieve the four concentrations of $1.00 \pm 0.06$ ppm, $2.0 \pm 0.1$, ppm, $5.0 \pm 0.3$ ,ppm, and $10.0 \pm 0.5$ ppm. The concentration of the xenon is determined from the volume and pressure of xenon added, and the measured weight of liquid argon in the cryostat.
The increase in uncertainty with dopant level is an accumulation of uncertainties in dopant as more xenon was added. Including the un-doped data then, we have a total of five data sets (referred to as sets 0,1,2,3, and 4 for the un-doped, 1 ppm, 2 ppm, 5 ppm, and 10 ppm respectively). 

Some data was taken in coincidence with scintillator paddles located above and below the cryostat as a cosmic trigger, however, all data presented here was self-triggered data.  Therefore, the events were caused by a mixture of cosmogenics and radiogenics.  

\section{Event selection and data set preparation}
\begin{figure}[htbp]
\begin{center}
 \includegraphics[width=0.9 \columnwidth]{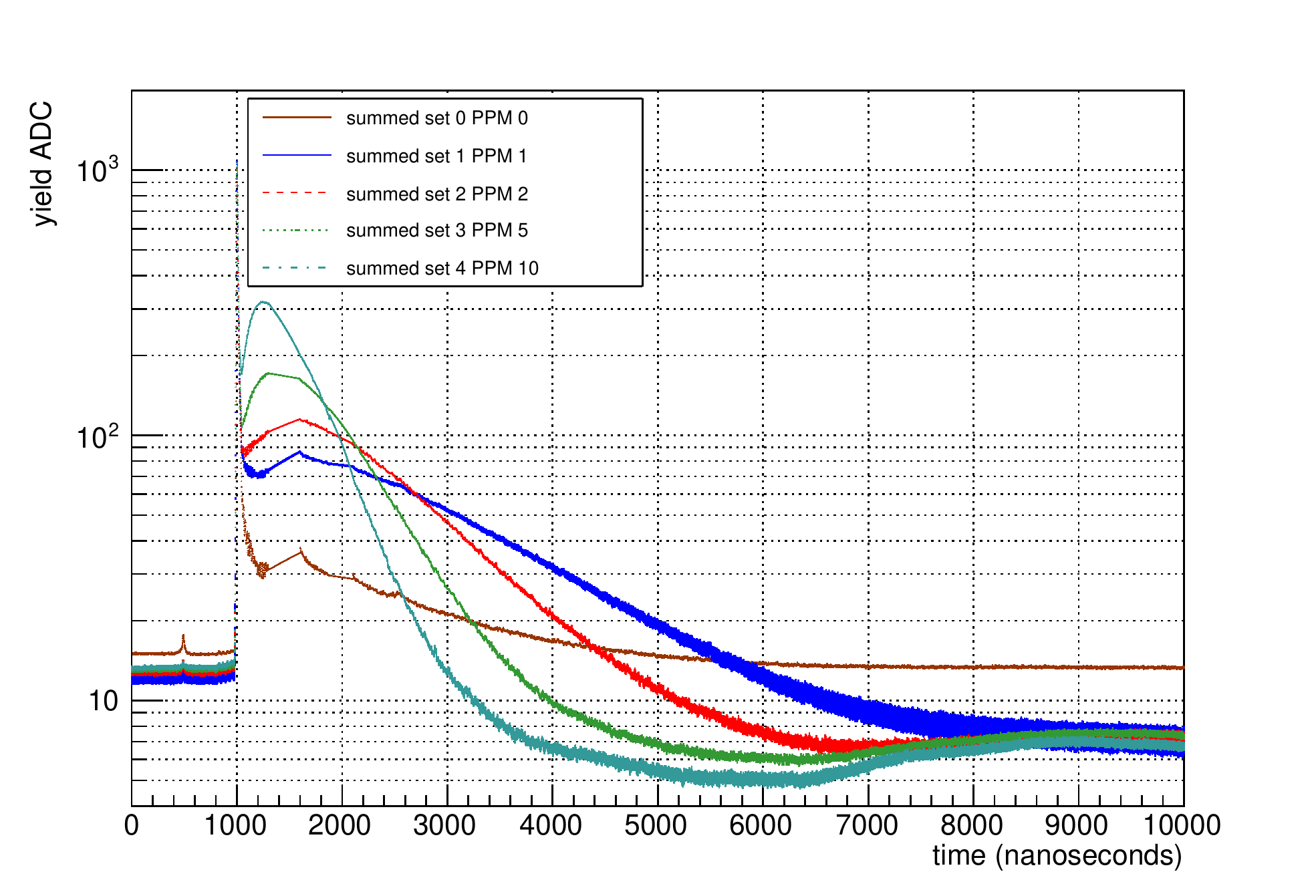}
\caption{
%{\color{red} Can we change the label to the dopant levels instead of data sets numbers that are arbitary, no x and ya xis label} 
Summed time distributions of the individual events as a function of doping level (0, 1, 2, 5, and 10). The baselines before the singlet peak are all centered around zero in the data, but a small offset has been added in order to plot the data on a log scale.}
\label{fig-all-summed}
\end{center}
\end{figure}

We modified the analysis algorithm used in Ref.~\cite{McFadden:2020dxs} due to concerns about after-pulsing in the PMT as well as concerns about the pulse-finding efficiency immediately following the large singlet pulse due to overshoot in the AC coupled PMT (a remnant of this overshoot can be seen in the first two data sets, see Fig. \ref{fig-all-summed}). The previous analysis used a derivative-based pulse-finding algorithm. %on individual waveforms.
%that may miss small light signals whose amplitude is close to the noise level. 
The new algorithm simply sums the event waveforms by run.  Additionally, we removed events using data clean-up cuts described below. Because we do not do pulse finding, we are unable to recover the small signal above the noise for late ($\rm >3000 \: ns$) times.
%in a doping run and makes the analysis more sensitive to small light emission, as they are present in the triplet light emission. 

 During data taking, individual $\rm 10 \:\mu s$ long waveforms are recorded. The data includes $1 \: \mu\rm s$ prior to the trigger from which a baseline is determined.  We used a threshold trigger tuned to $\sim 15$ PE. This value was estimated by comparing the singlet peak of the waveforms to the individual small amplitude pulses in the late light. This threshold was introduced to reduce noise triggers while still remaining very efficient for ionization events. With this threshold, our data was virtually all due to cosmic rays. However, it should be noted for later reference that triggered events represent ionizing events from all orientations and distances to the PMT, and because of the effects of xenon doping, the subset of these cosmic rays may differ between dopant sets.
 
\begin{figure}[htbp]
\begin{center}
 \includegraphics[width=0.9 \columnwidth]{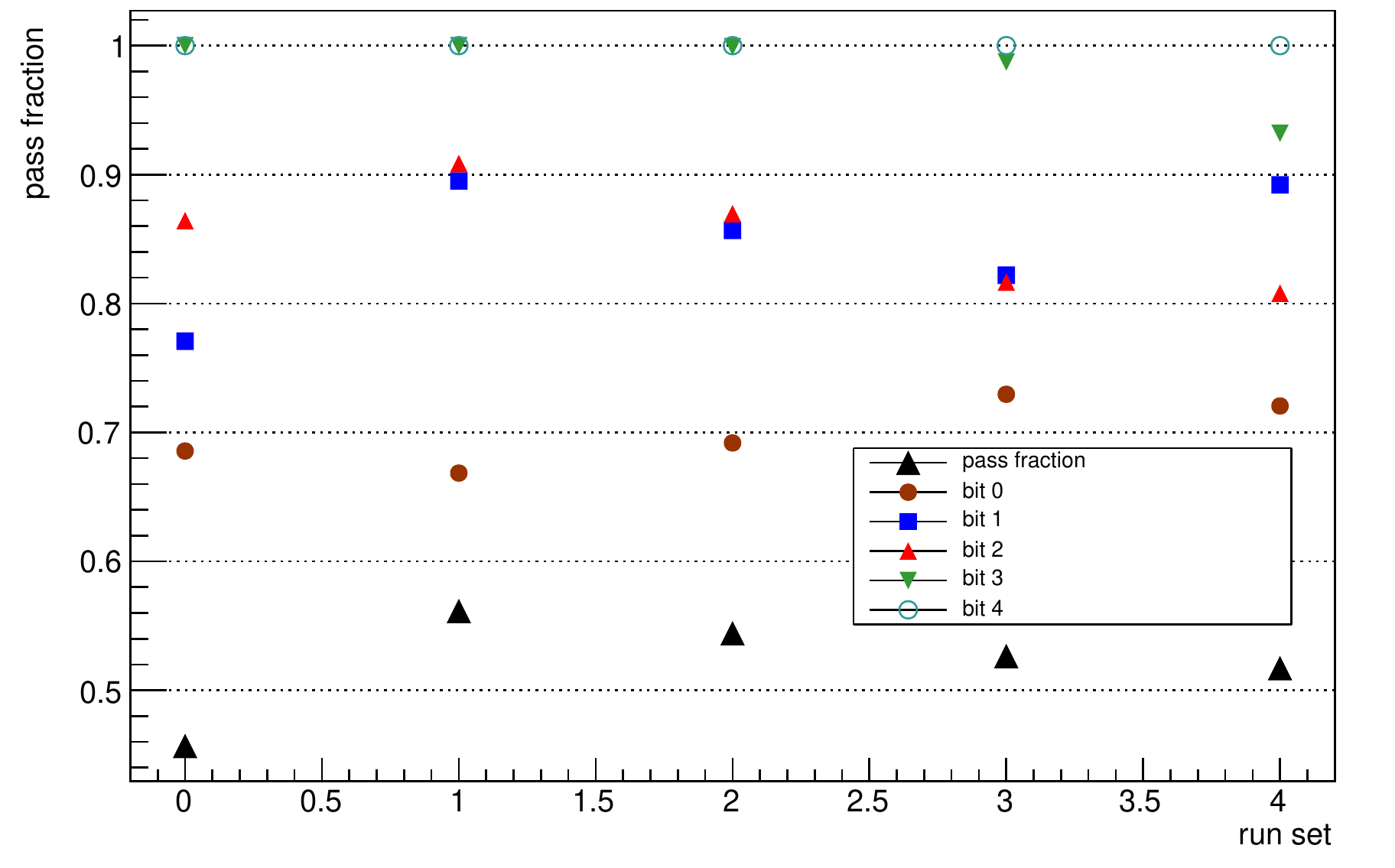}
\caption{Event clean-up cuts for each data set with cut bits are defined in the text.
The sets 0-4 correspond to the doping by 0,1,2,5, and 10 ppm xenon.}
%{\color{red} Can you change the label to data set and give it the 0,1,2,5,10 ppm lable ?}}
\label{fig-pass}
\end{center}
\end{figure}

\begin{figure}[htbp]
\centering
 \includegraphics[width=0.9 \columnwidth]{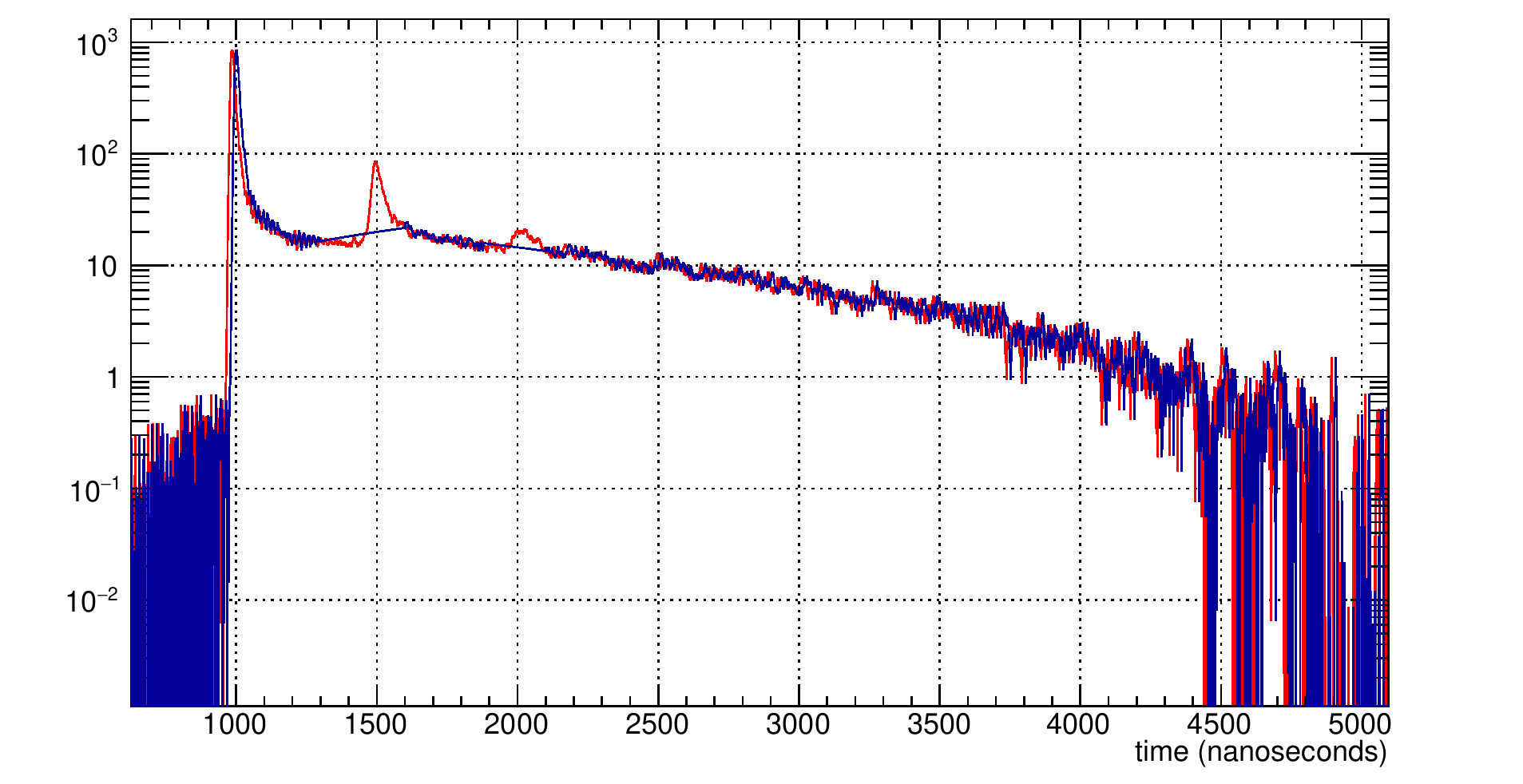}
\caption{
%{\color{red} This is a run not one event. can you remove the title} 
Typical time distribution in an event showing PMT after-pulsing (red). In this case, the event is from the un-doped data set. Each event has the after-pulsing removed by interpolation (blue). The peak time for each event is also shifted to align at 1000 ns.}
\label{fig-vb}
\end{figure}

 A series of cuts was applied to remove saturated and pile-up events (labeled as bits 0-4 in Figure~\ref{fig-pass}). Saturated events, which come from high energy cosmic ray events, create a false singlet to non-singlet light ratio and are removed (bit 0). Individual waveforms which show either a pulse before the singlet peak (bit 1) or mismatch in baselines at the beginning and the end of the event (bit 2) are also removed from the data set. These cases do not allow for a correct estimate of the singlet value due to a disturbed baseline. Pile-up events and events with high noise can be identified by the presence of a second large peak after the singlet (bit 3). figure~\ref{fig-pass} shows the overall acceptance is $\sim 55\%$ and approximately constant across dopant sets. Individual waveforms have their baseline subtracted, and the peak of the singlet is aligned to $t=$1000\,ns to avoid additional smearing due to these time offsets. Additionally, a PMT after-pulsing artifact (due to presence of helium in the PMT) is removed by a simple linear interpolation algorithm (see figure~\ref{fig-vb}.)  All events which pass the data quality cuts are then summed by run.
 
\begin{figure}[htbp]
\begin{center}
\includegraphics[width=0.9\columnwidth]{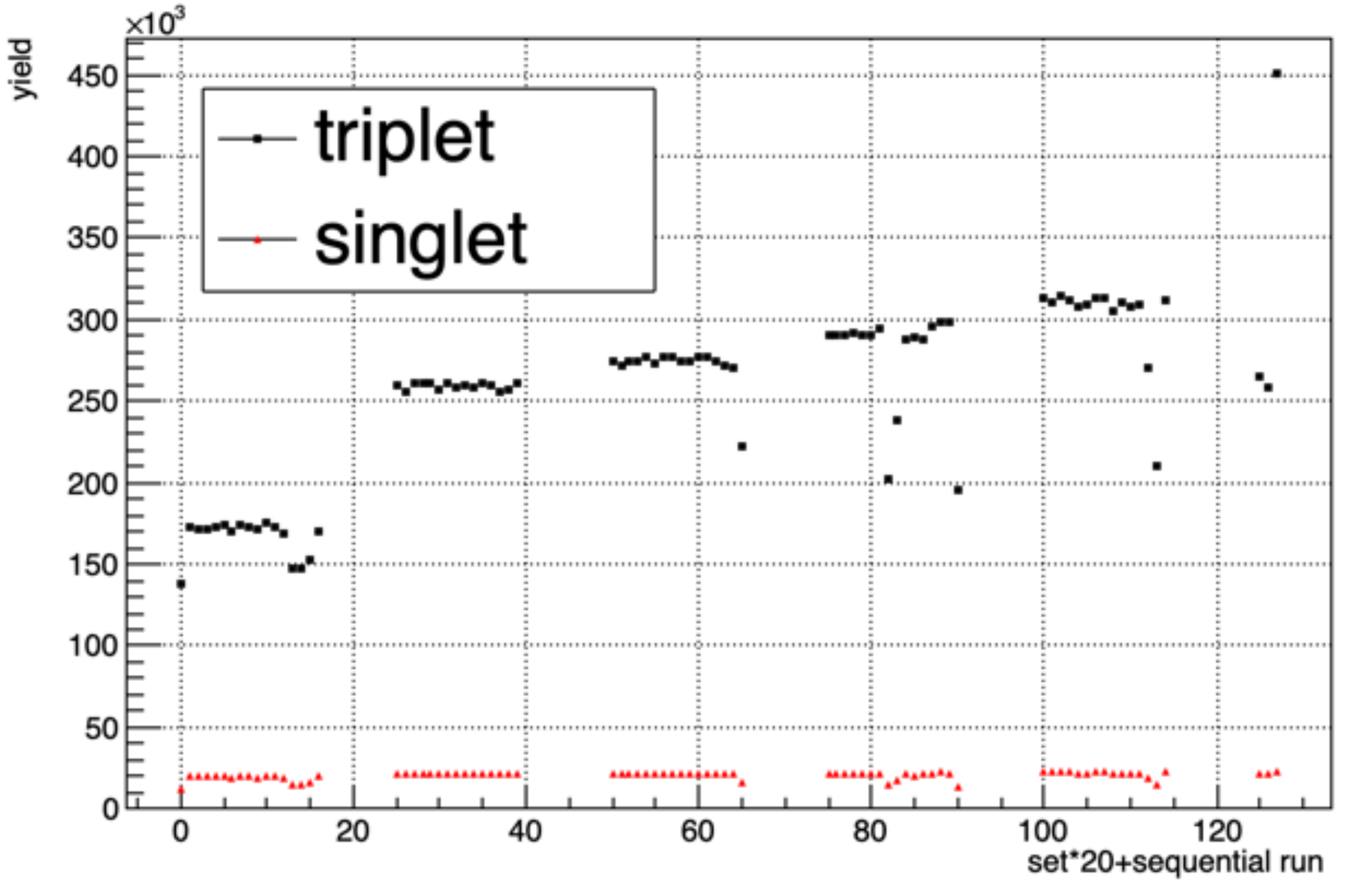}
\caption{Relative light yield in the singlet (summed from 900 to 1020 ns) and triplet region (summed from 1020 to 2300 ns) by run number to monitor the stability of the doping.}
\label{fig-by-run}
\end{center}
\end{figure}

\begin{figure}
\begin{center}
 \includegraphics[width=0.9 \columnwidth]{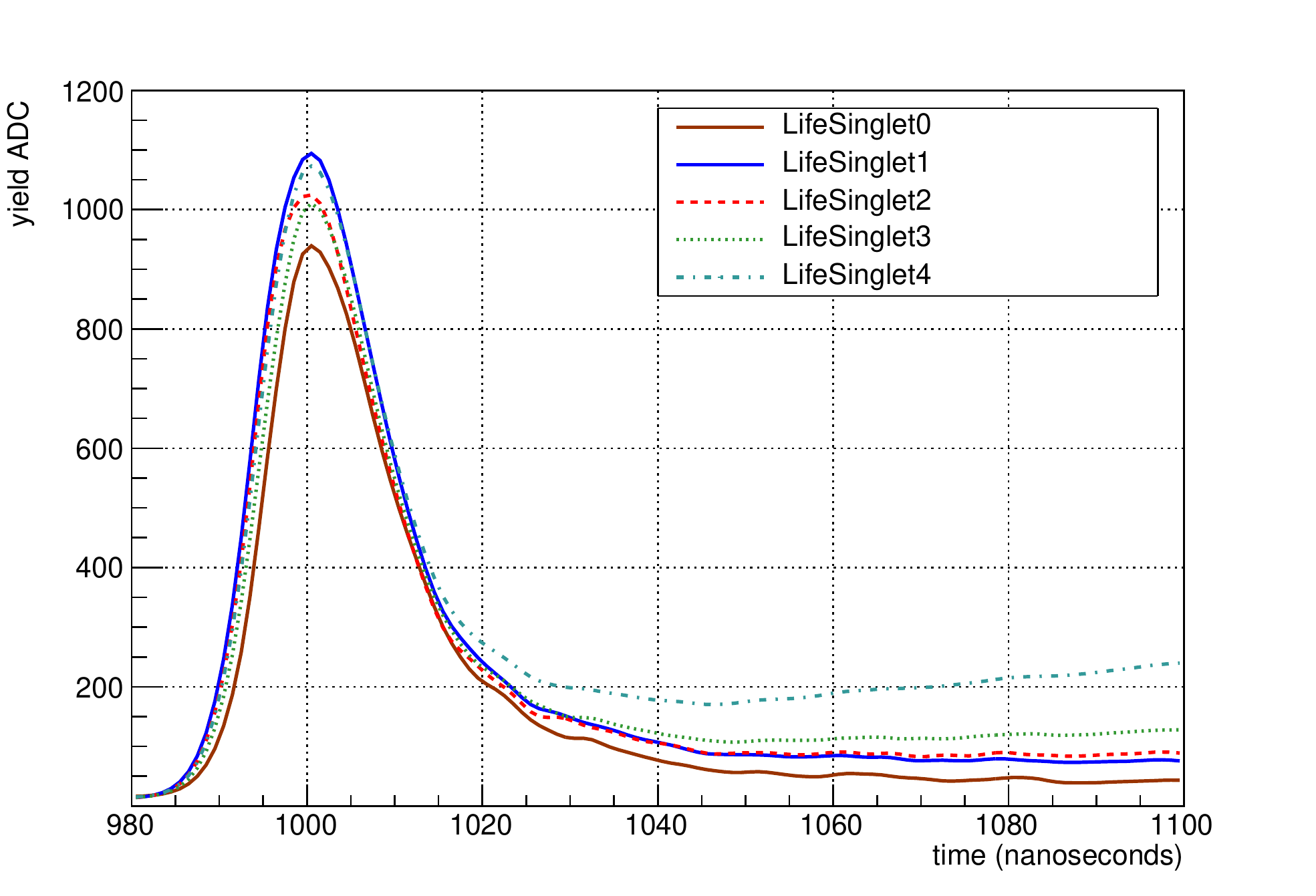}
\caption{
%{\color{red} Can we change the label to the dopant levels instead of data sets numbers that are arbitrary, whats SPE and why 40ns ? } 
Zoom in into the singlet region of figure~\ref{fig-all-summed} with the same color coding. The rise time of the singlet peak (approximately 6 ns) was used to determine the timing resolution of the setup. }
\label{fig-singlet-summed}
\end{center}
\end{figure}

\begin{figure}[htbp]
\begin{center}
 \includegraphics[width=0.9 \columnwidth]{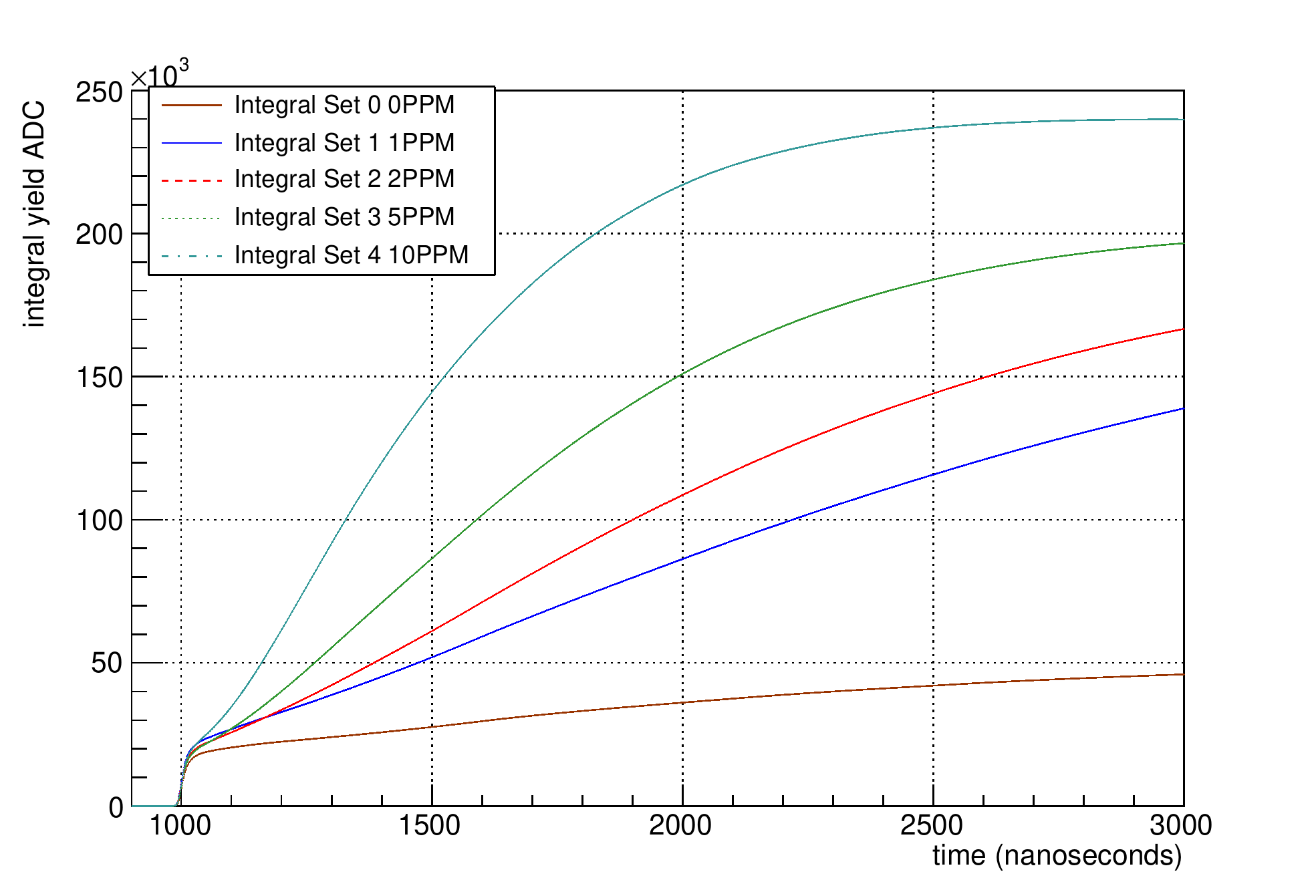}
\caption { 
%{\color{red} Do we need the integrated distribution s Can we change the label to the dopant levels instead of this cryptic name ?, whats SPE and 40ns} 
Integrated time distributions  of Figure\,\ref{fig-all-summed} (1,000 to 3000 ns) for each with doping level 0, 1, 2, 5, and 10. The shift in the light to early times as well as the increase in light with dopant is apparent.}
\label{fig-all-integral}
\end{center}
\end{figure}

 The integrated light from the singlet peak (900-1020 ns) and the late light region (1020-2030 ns) is shown run-by-run in figure~\ref{fig-by-run}. The late light region is defined to avoid late-time baseline shifts that we do not attempt to correct (see figure~\ref{fig-all-summed}). A step-wise increase in the late light region with dopant is clearly apparent, with sets corresponding to  0,1,2,5 and 10 ppm, whereas the singlet light changes very little with dopant (see figure~\ref{fig-singlet-summed}). A few obvious outlying runs are removed by hand in the further analysis.  The waveforms are then summed by dopant sets and normalized to the number of events in each set (See figure~\ref{fig-all-summed}).
  In total about 400000 events for each doping set allow an analysis of the light time distribution with negligible statistical uncertainty. The increase of light with dopant as well as the shift of the light to earlier times is apparent
in the integral waveforms shown in figure~\ref{fig-all-integral}.
 %Each doping level consists about 20ish runs, see Fig. \ref{fig-by-run}.  ({\color{red} how long is one run and each data set recorded, and are their any mix-in times we shuold consider}).
 %Fig. \ref{fig-vb} shows one of these run-subsets from the un-doped data. While not observable on the each individual waveform, the sum reveals that after pulsing is not negligible. The two after pulsing regions 500-ns and 1000-ns after the trigger have been removed by linearly interpolating between regions before and after. 
%Fig.~\ref{fig-by-run} shows how the amount of light in the singlet region (900 to 1020 ns) and triplet region (1020 to 10000 ns ) varies as a function of doping. Since each event is normalized to the amount of light in the singlet, it is constant and a verification of the stability of the system and the analysis. Only a few runs showed increased noise and have therefore a slightly smaller amount of recorded events. These run ranges have been removed from the analysis as well.  The pattern of increasing triplet light yield with dopant is apparent. When combining all run subsets and normalizing them by the number of evens, one receives the time distribution as a function of doping, see Fig.~\ref{fig-all-summed} and Fig.~\ref{fig-singlet-summed}. We have tried different baseline correction methods, however the uncertainty introduced by these methods was larger than the correction needed since the sagging is small compared to the statistics in the singlet, after-singlet, and triplet regions we plan to analyze, see also Fig.~\ref{fig-all-integral}.

 \section{Model fitting results} 
 
\begin{figure}[htbp]
\begin{center}
 \includegraphics[width=0.45 \columnwidth]{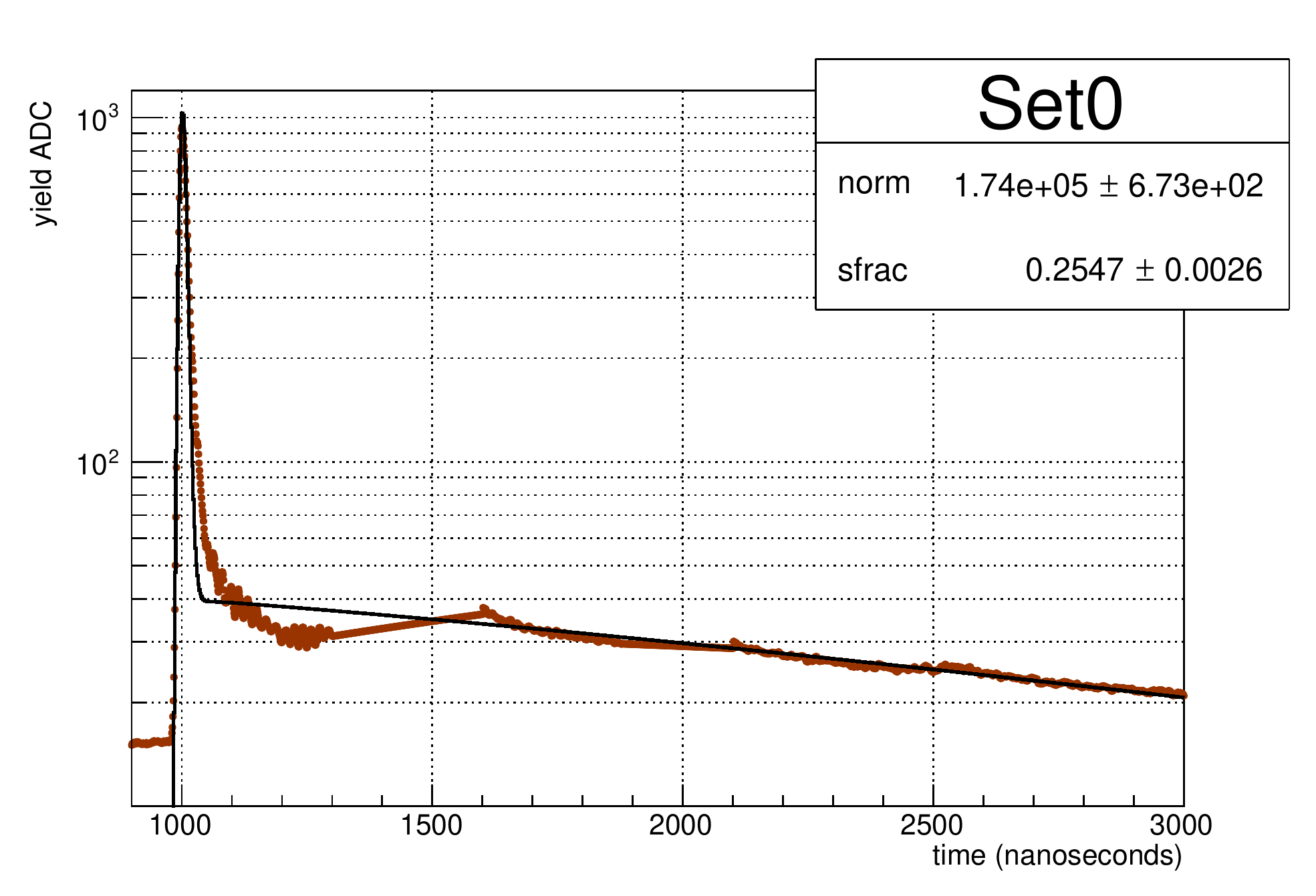} 
  \includegraphics[width=0.45 \columnwidth]{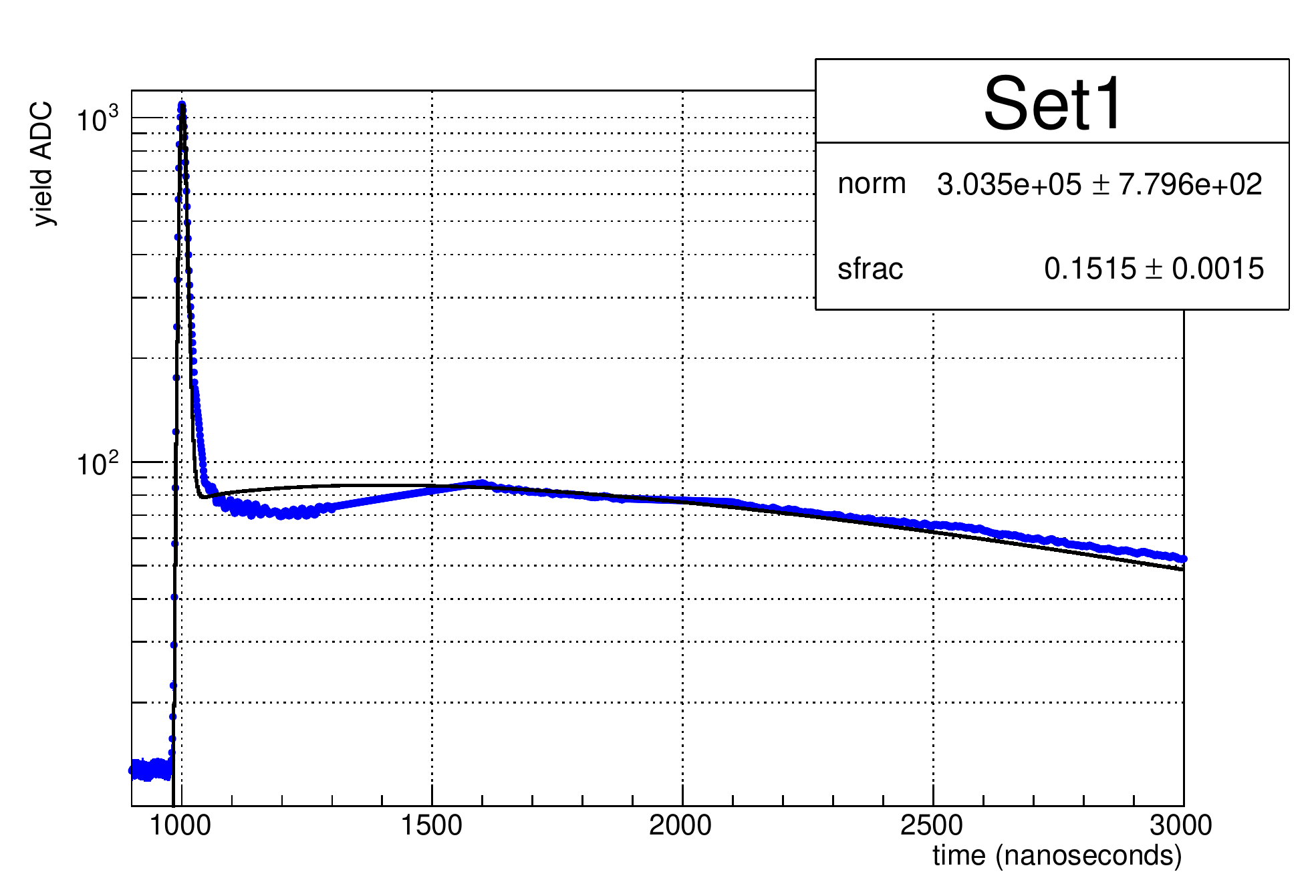}
  \includegraphics[width=0.45 \columnwidth]{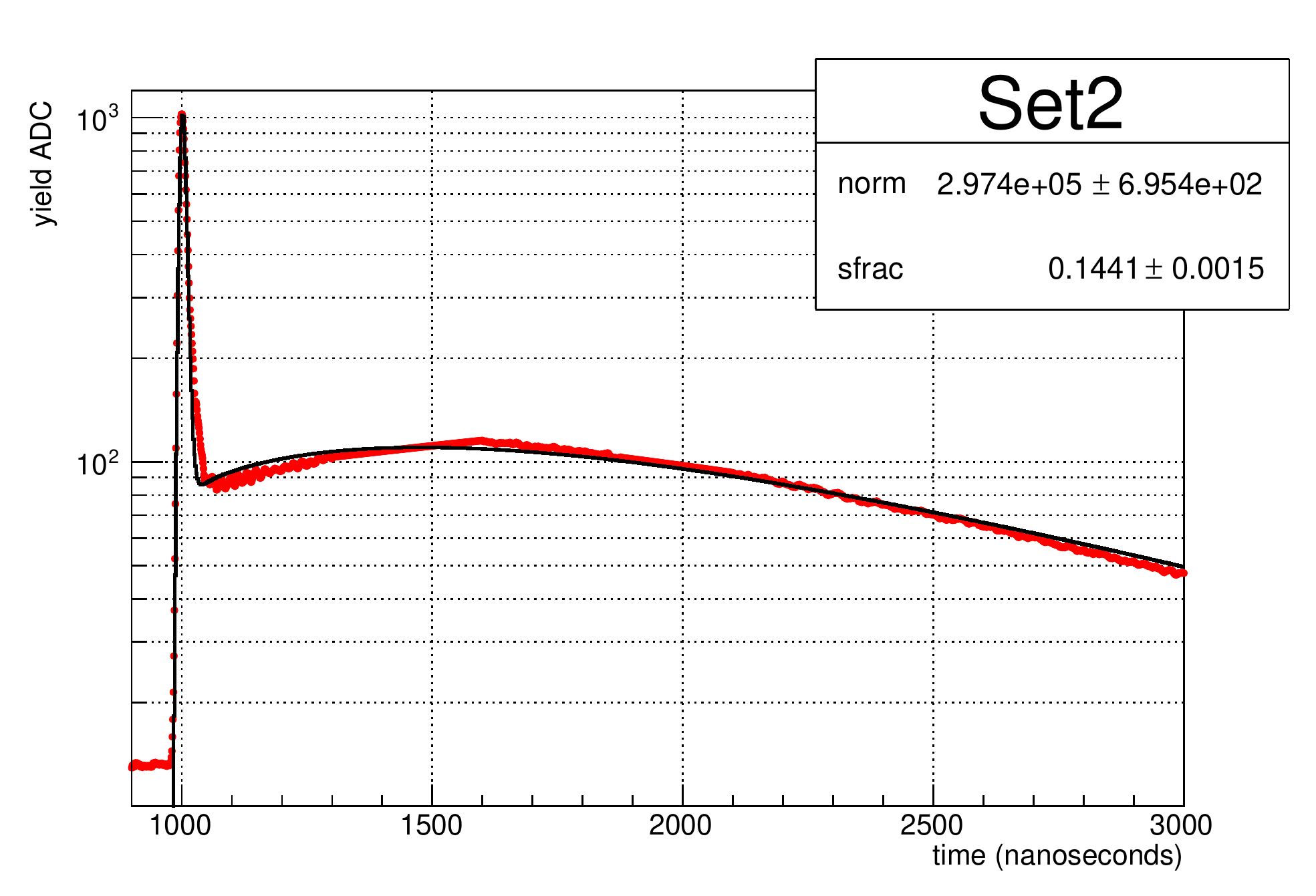}
  \includegraphics[width=0.45 \columnwidth]{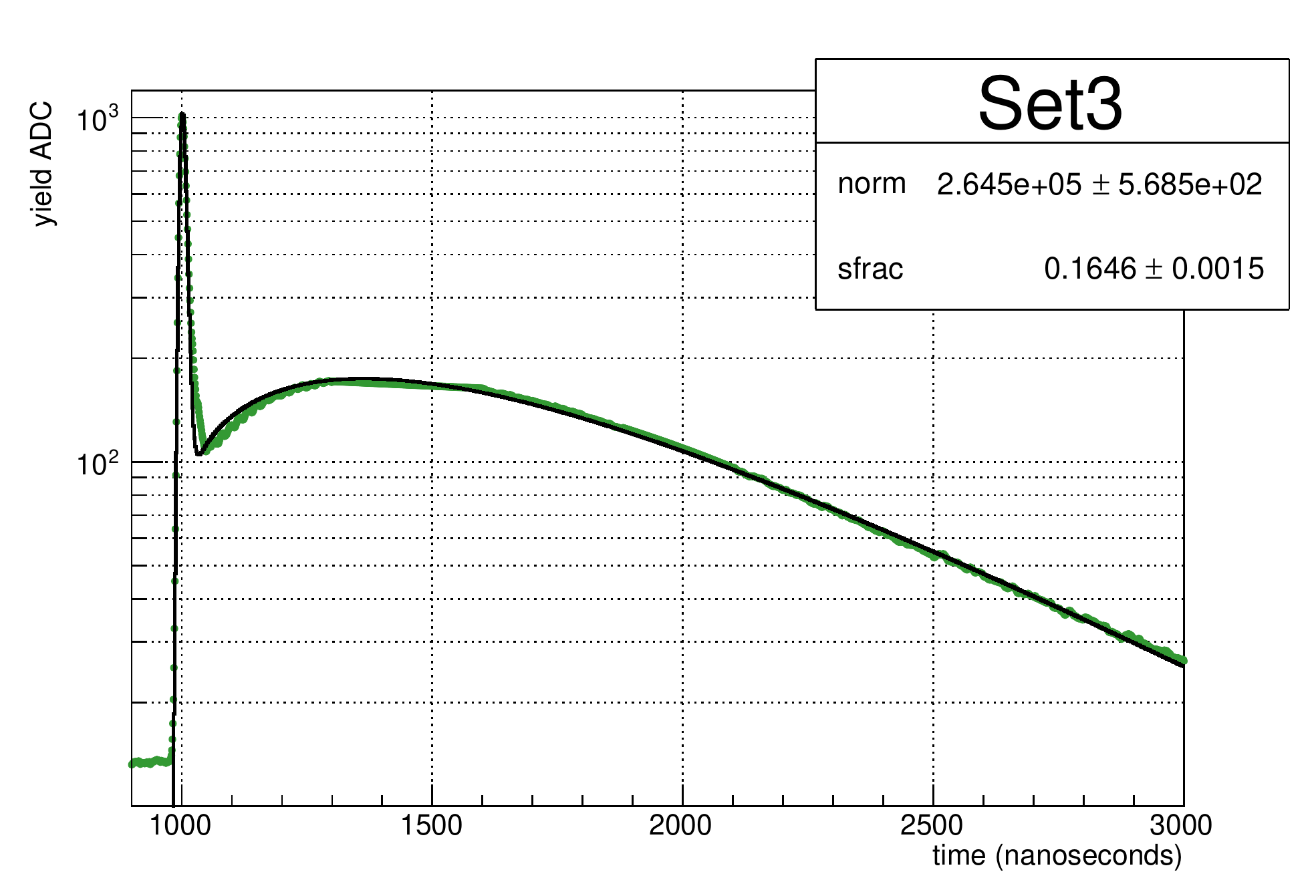}
  \includegraphics[width=0.45 \columnwidth]{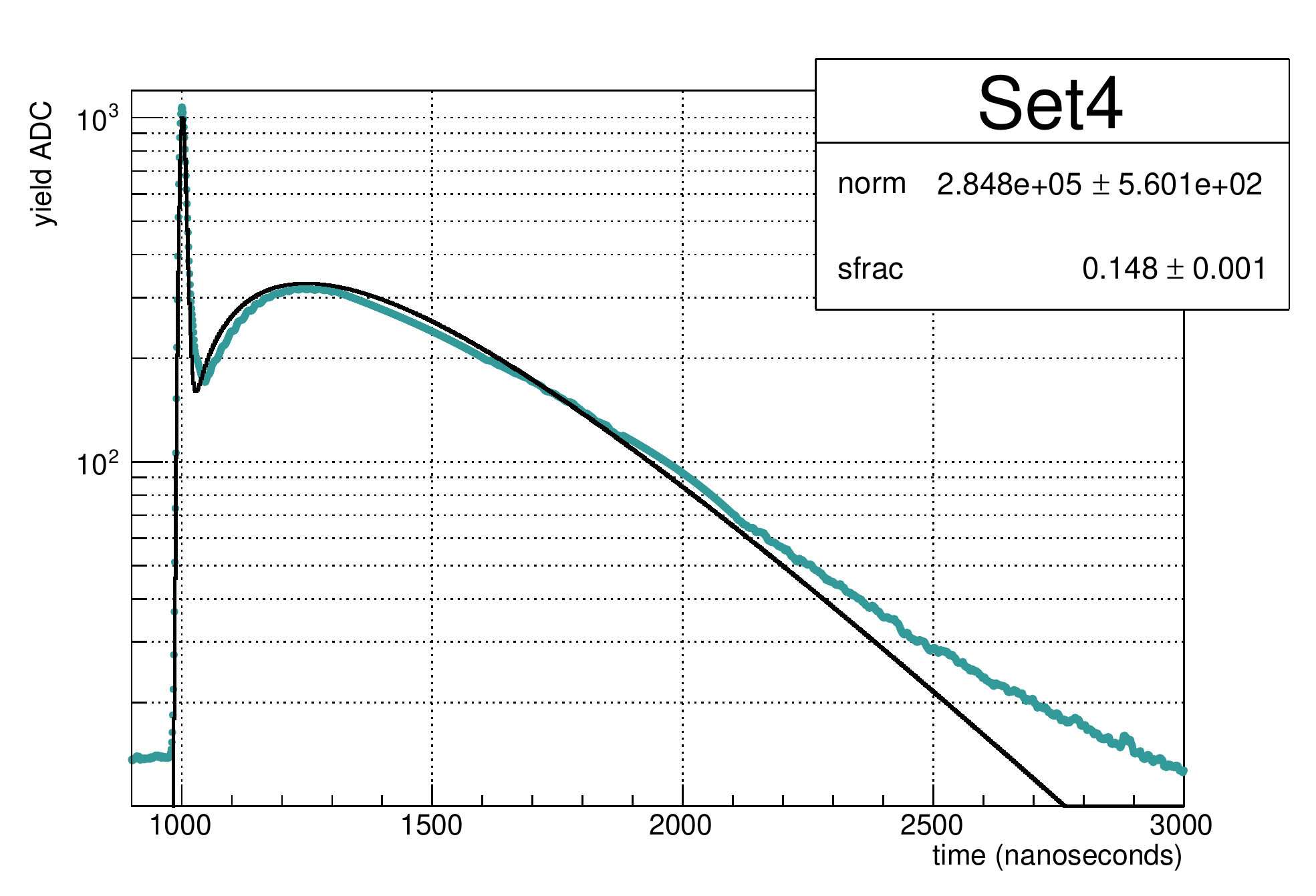}
 \caption{ 
 %{\color{red} can you give them better stat box labels with the dopant level and remove the titles ?. or remove the stat boxes at all and give them just the dopant as title} 
 Fits to the different doping data sets. The solid curve fits are the model with just two fit parameters (norm and sfrac, as described in the text).  The xenon concentrations are shifted by 0.4 relative to the added xenon, representing the residual xenon in our (commercially available, extracted from air) argon, and has a better fit to the undoped data. }
\label{figfit}
\end{center}
\end{figure}

 The time distribution of detected light from each set are fitted with the model for each level of doping with only two free parameters for each data set - the normalization (norm) and the fraction of singlet to triplet states (sfrac).  The fit region was limited to the region between the trigger time of 980 ns and 3000 ns
 to avoid problems with the baseline and since the biggest impact of doping is seen in this region.  Since we don't have a measurement of the residual xenon in our undoped argon, we fit by hand the xenon concentration for that set, and used the resulting value (0.4 ppm) as an offset for the remaining sets. With this offset, the constant value of absorbtion, $A=0.62$, used for all data sets is justified.  The fits are shown in figure~\ref{figfit}. All curves show good agreement with the measured light distribution in each data set. 
 
The normalization can be thought of as the total number of initial singlet plus triplet argon excimers (after correction for acceptance and efficiency), and should not depend on doping levels.  However, since the absorption of the 128nm light can affect the trigger bias, the relatively small variations seen in the normalization are not considered significant.

 %Fig.~\ref{figppm} shows how the fitted amount of doping agrees well with the known doping level. 

\section{Discussion}

\begin{figure}[htbp]
\begin{center}
\includegraphics[width=.9 \columnwidth]{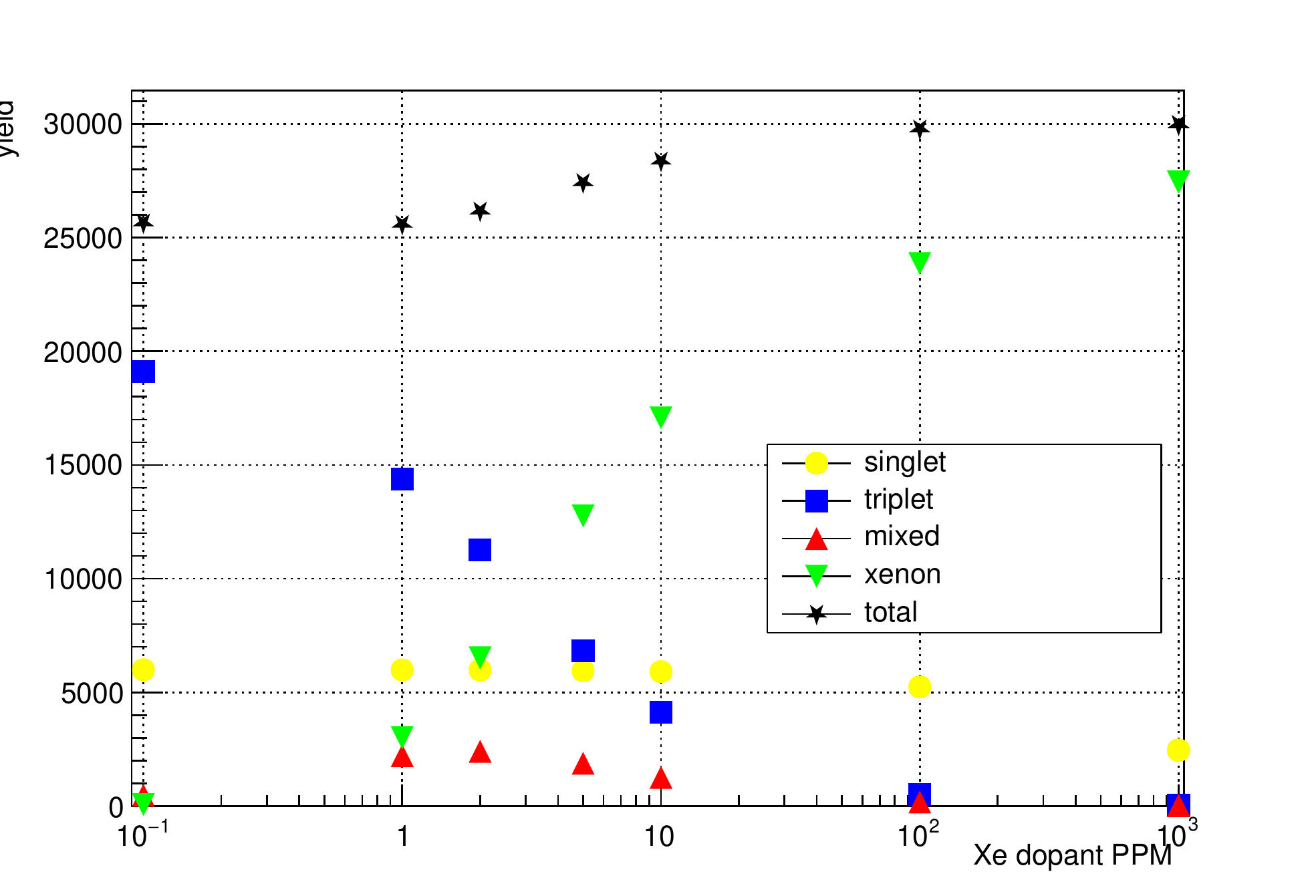}

\caption{Model light yield prediction with   $A=0$ for all xenon concentrations. We set the total number of initial excimers to be 30,000 and integrated from the trigger at 1000 ns to 10000 ns. Markers are light from singlet (yellow dots), triplet (blue squares), mixed (red triangles), xenon (green inverted triangles) and total (black stars).}
\label{fig-model}
\end{center}
\end{figure}

\begin{figure}[htbp]
\begin{center}
\includegraphics[width=.9 \columnwidth]{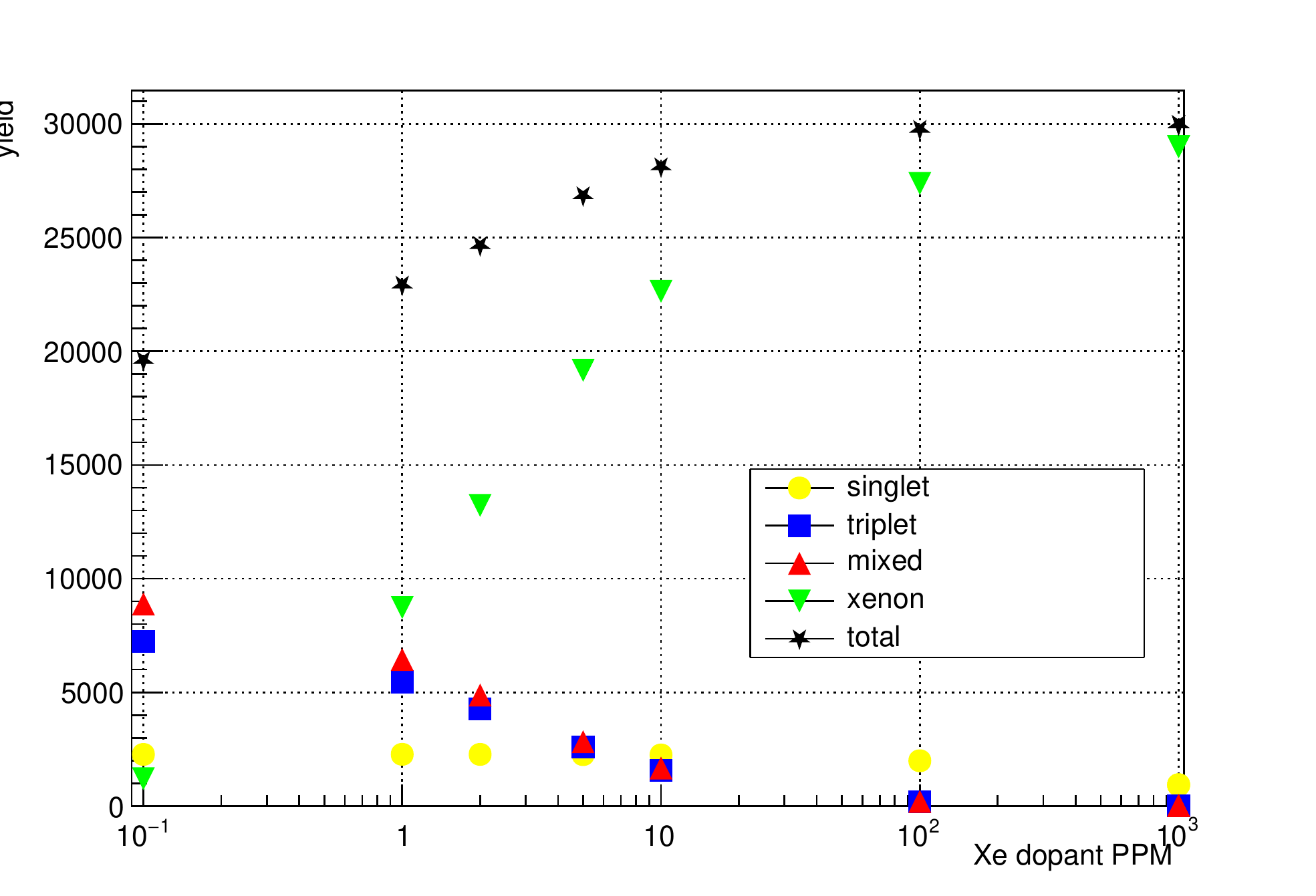}
\caption{Model light yield prediction with $A=0.62$ kept as a constant over xenon concentrations. We set the total number of initial excimers to be 30,000 and integrated from the trigger at 1000 ns to 10000 ns. Markers are light from singlet (yellow dots), triplet (blue squares), mixed (red triangles), xenon (green inverted triangles) and total (black stars).}
\label{fig-model-two}
\end{center}
\end{figure}

%Based on the good agreement between measured time distributions and the prediction of the model, we studied the influence of the absorption parameter.
We can use our model to predict the light produced by the separate excimers as a function of dopant. These curves are shown in  figure~\ref{fig-model} and figure~\ref{fig-model-two}, which shows the time-integrated light yield of the individual components as a function of doping. If the absorption is set to zero (figure~\ref{fig-model} ) the results are in good agreement with the model by Segreto \cite{Segreto:2020qks} which indicates a 20\% increase in total light yield from 0 ppm to 100 ppm xenon concentration, where we see a 15\% increase from 0.1 ppm to 100 ppm without absorption. In our model (without absorption) this increase is due to the transfer of the argon excimer state to the mixed state with dopant, a process which competes with the quenching of the argon state. However, the model of that reference does not include  absorption by atomic xenon which increases the rate of mixed state formation.  Furthermore, that model doesn't include quenching of the mixed state. 
With the inclusion of absorption ( figure~~\ref{fig-model-two}), results from our data and the results from Ref.\cite{Neumeier:2015lka} are well described. 
%where a large increase between un-doped argon and doped argon can be described well. With absorption, 
The light yield without dopant is greatly suppressed by the large absorption and creation of mixed excimers with a long lifetime.
Both the triplet excimers and the mixed excimers are quenched with the result that much of the light is lost.
Addition of the xenon dopant increases the rate of transfer to the xenon excimer state and recovers this lost light.
%from the argon triplet and the mixed excimer states.
%many of which subsequently get quenched. Hence, while the mechanism of light loss at low xenon dopant levels is the same (collisional quenching of the mixed and triplet states), absorption of the 128 nm emission greatly enhances this effect at those low levels.

\begin{figure}[htbp]
\begin{center}
       \includegraphics[width=.9 \columnwidth]{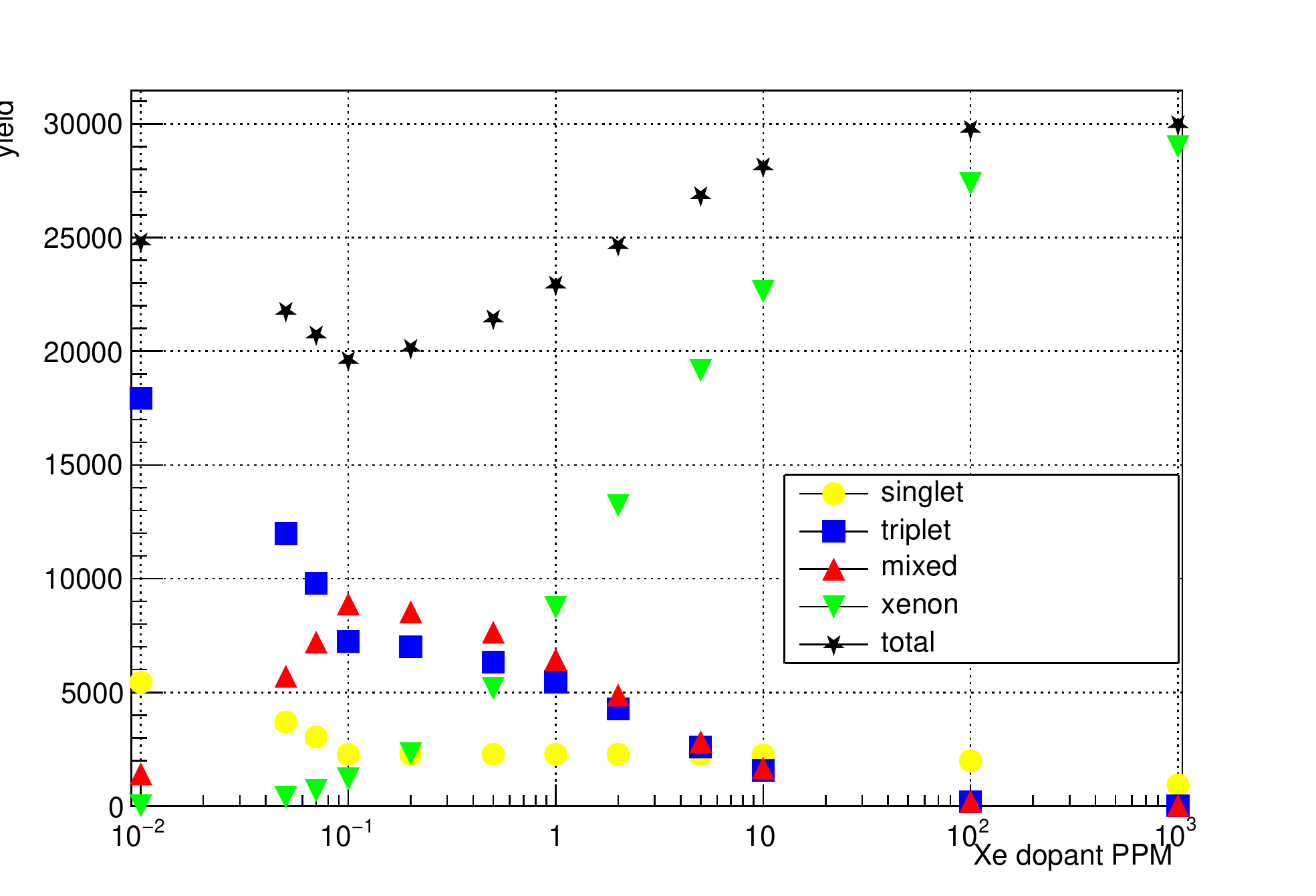}
\caption{Model light yield prediction with $A$ taken as a function of xenon concentration based on the double exponential fit of figure~\ref{fig-attVdist}. 
We set the total number of initial excimers to be 30,000 and integrated from the trigger at 1000 ns to 10000 ns. Markers are light from singlet (yellow dots), triplet (blue squares), mixed (red triangles), xenon (green inverted triangles) and total (black stars).}
\label{fig-model-low}
\end{center}
\end{figure}

To use this model at low xenon concentrations ($< 0.1$ ppm), we can use the curve of figure~\ref{fig-attVdist} to determine the absorption factor corresponding to these low dopant levels.  In this case the absorption decreases linearly from the constant A=0.62 to zero.  Thus, at these very low xenon concentrations the light yield also increases (see figure~\ref{fig-model-low}).

%one would have to use the convoluted data of figure~\ref{fig-attVdist} to determithne the absorption at these low dopant levels.  Doing that, one gets the model results shown in Fig. \ref{fig-model-low}, which demonstrates that the light yield \textit{increase} seen is, as expected, just a light yield \textit{recovery} from light yield loss due to trace amounts of xenon (0.001 - 1 ppm) found in commercial argon. 

Finally, there has been some concern with xenon doping that particle identification using the ratio of fast to slow light would be negatively impacted.  However, as seen from our model, the singlet light yield is not significantly impacted at $\sim$ 10 ppm relative to the residual level of 0.4\.ppm, and while there is more light at earlier times (but after the singlet peak), it should not impact the uncertainty in the ratio of the ``fast'' to ``slow'' light, only the value of the ratio.  This ratio for different primary ionization will still be determined by the relative strength of the initial singlet light, however, a tighter time window on the singlet peak will be necessary.

\section{Summary and Outlook}
In this work we presented a model for light produced by xenon doping that accurately describes the time development of the light as measured in our experiment.  The increased light yield by a factor of $ 1.92 \pm 0.12$ at 10 ppm over the yield in argon with traces of Xenon up to a few ppm as reported in Ref.~\cite{McFadden:2020dxs} is now understood as the release of the excimer trapped light with the addition of xenon.
Thus, the addition of 10 ppm xenon to liquid argon detectors increases the light yield in our geometry and trigger set by a factor of two, as well as shifting the light to early times and to the more favorable 175 nm wavelength. 

In future work, we will confirm our model by measuring the separate (128 and 175 nm) emission wavelengths as a function of distance and emission time.  Additionally, by using a triggered source (design provided by Sch\"{o}nert's group at TUM\cite{PrivateTUM}), by doping in finer steps below 1 ppm and by measuring the light yield accurately out to 10 $\mu$s we expect to extract values for the quenching rate and the absolute light yield in PE per deposited ionization.

\section{Acknowledgements}
We gratefully acknowledge the support of the U.S. Department of Energy through the LANL/LDRD Program and by the U.S. Department of Energy, Office of Science, Office of Nuclear Physics award LANL-EM78.
%We gratefully acknowledge the support of...
%that describes the increase of observed light over a wide range of doping. The model is able to describe observed time distribution of emitted light in our test stand. With a hardware upgrade currently in progress, we plan to measure the absolute light yield using a calibrated source and separate the measured light 
%emission by wavelength. By using an array of detectors with different sensitivities we plan to verify the predicted light emission from xenon and argon.  Given that xenon doping shifts the light to a more favorable wavelength as well as make the light more prompt, the exact determination of light emitted at this wavelength will help to design future experimental efforts. 

%{\color{red} do we need this discussion of where the light yield come from ?}
%The accepted maximum value for the scintillation yield in pure liquid argon is $\rm 50/keV_{ee}$ 
%($\rm40/keV_{ee}$ for electrons).\cite{Doke:2002oab}.
%This value is not a measured yield, but rather comes from two related experimental numbers, the ratio of 
%excited to ionized argon atoms $N_{ex}/N_i = 0.2$
%and the ionization work function $W=23.6\ eV.$\cite{Doke:2002oab},\cite{Crawford:1987qq},\cite{Foreman:2019dzm} Then the maximum yield in photons per keV is 
%$\rm ( 1+N_{ex}/N_i )/ W = 50 \; photons/keV .$

%An additional benefit of transferring the scintillation to xenon excimers is the higher reflectivity and 
%longer attenuation length of the $175$ nm light in liquid argon.

%-----------------------------------------------------------
\clearpage

\bibliographystyle{elsarticle-num-names}
\bibliography{biblio}

\clearpage

\end{document}